\documentclass{iaus}
\usepackage{graphicx}
\usepackage{lipsum}
\usepackage{natbib}
\usepackage{amssymb}

\title[Modelling of an eclipsing RS\,CVn-binary: V405\,And] 
{Modelling of an eclipsing RS\,CVn-binary: V405\,And}

\author[K. Vida, K. Ol\'ah, Zs. K\H{o}v\'ari]   
{Kriszti\'an Vida$^1$,
Katalin Ol\'ah$^1$,
Zsolt K\H{o}v\'ari$^1$}

\affiliation{
$^1$ Konkoly Observatory of the Hungarian Academy of Sciences, \\ 
H-1121 Budapest, Konkoly Thege Mikl\'os \'ut 15-17, Hungary
}

\pubyear{2011}
\volume{282}  
\pagerange{119--126}
\jname{From Interacting Binaries to Exoplanets:
Essential Modeling Tools}
\editors{A.C. Editor, B.D. Editor \& C.E. Editor, eds.}
\begin{document}

\maketitle

\begin{abstract}
V405\,And is an ultrafast-rotating ($P_\mathrm{rot}\approx0.46$ days) eclipsing binary.
The system consists of a primary star with radiative core and convective envelope, and a fully convective secondary.
Theories have shown, that stellar structure can depend on magnetic activity, i.e.,
magnetically active M-dwarfs should have larger radii. Earlier
light curve modelling of V405\,And indeed showed this behaviour:
we found that the radius of the primary is significantly larger than
the theoretically predicted value for inactive main sequence stars (the discrepancy is the largest
of all known objects), while the secondary fits well to the
mass-radius relation.
By modelling our recently obtained light curves, which show significant changes of the spotted
surface of the primary, we can find further proof for this phenomenon.
\keywords{stars:activity, binaries: eclipsing, stars: fundamental parameters, stars: late-type, stars: spots}
\end{abstract}

\firstsection 
\section{Introduction}
V405\,And is an X-ray emitting active binary detected by the ROSAT satellite \citep{rosat}. The first detailed study of the system was done by \cite{chil}, who detected an orbital period of $P_\mathrm{orb}=0.465$ days, and a small, near grazing eclipse. The authors found that the primary and the secondary have spectral types of M0V and M5V, and both of them are active, as both show H$\alpha$ emission. 
\cite{vand} presented photometric $BV(RI)_C$ data, analysed optical spectroscopic measurements, and found, that the light curve modulation is caused by the combined effect of spottedness and binarity. Using an iterative modelling method to separately describe these two effects, the authors determined the physical properties of this binary system. The primary and the secondary component was found to have masses of $0.49\,M_\odot$ and $0.21\,M_\odot$ respectively, thus  the primary is supposed to consist of a radiative core, and a convective envelope, while the secondary is probably fully convective. The radii of the two components are $0.78\,R_\odot$ and $0.24\,R_\odot$. Plotting these values on the theoretical mass-radius diagram of \cite{baraffe} together with other binaries, we find that the secondary fits well to this relation, but the primary has a significantly larger radius than the theoretically predicted value.

\section{Observations and analysis}
\begin{figure}[t]
\begin{center}
\includegraphics[angle=-90,width=0.49\textwidth]{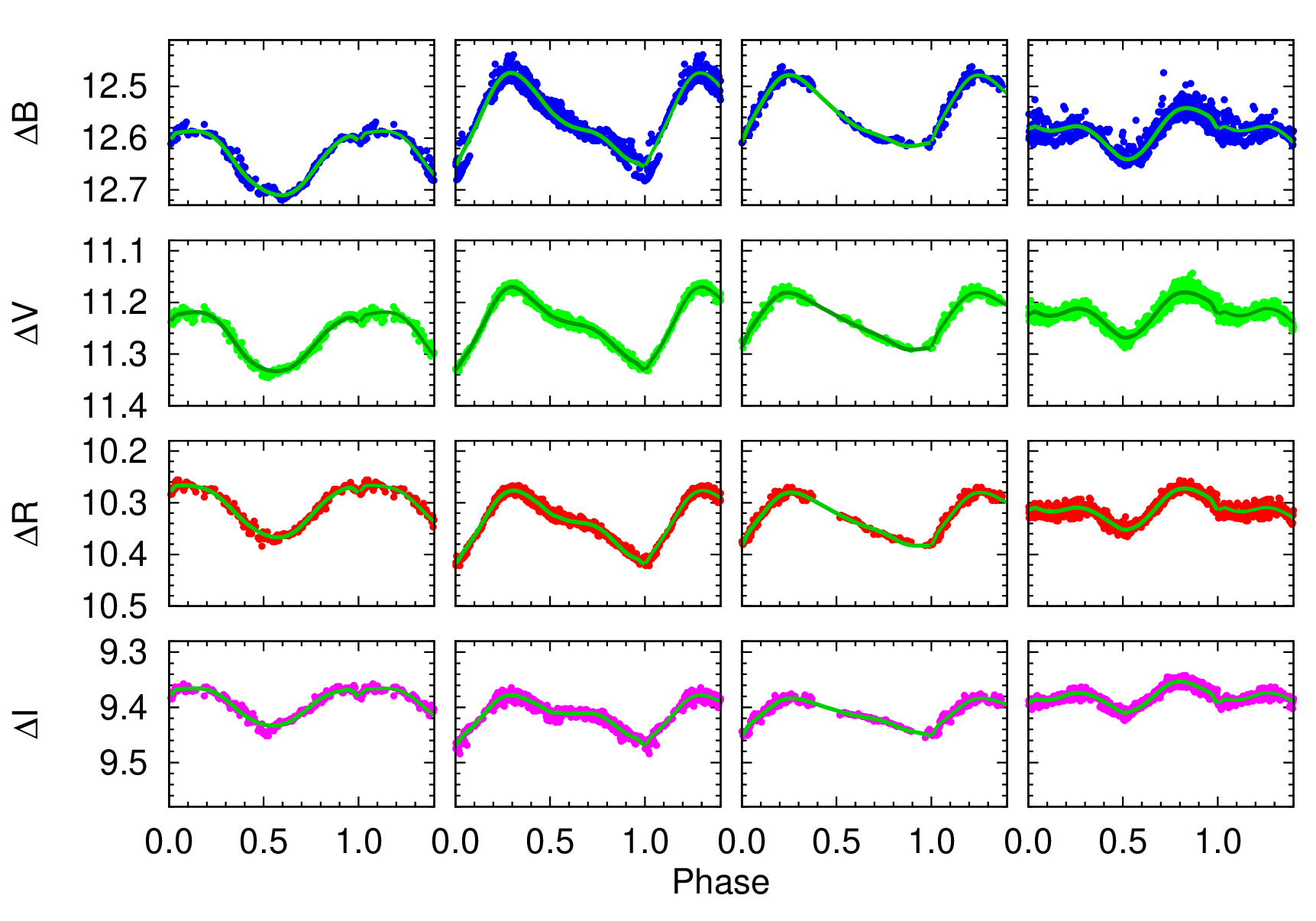} 
\includegraphics[angle=-90,width=0.49\textwidth]{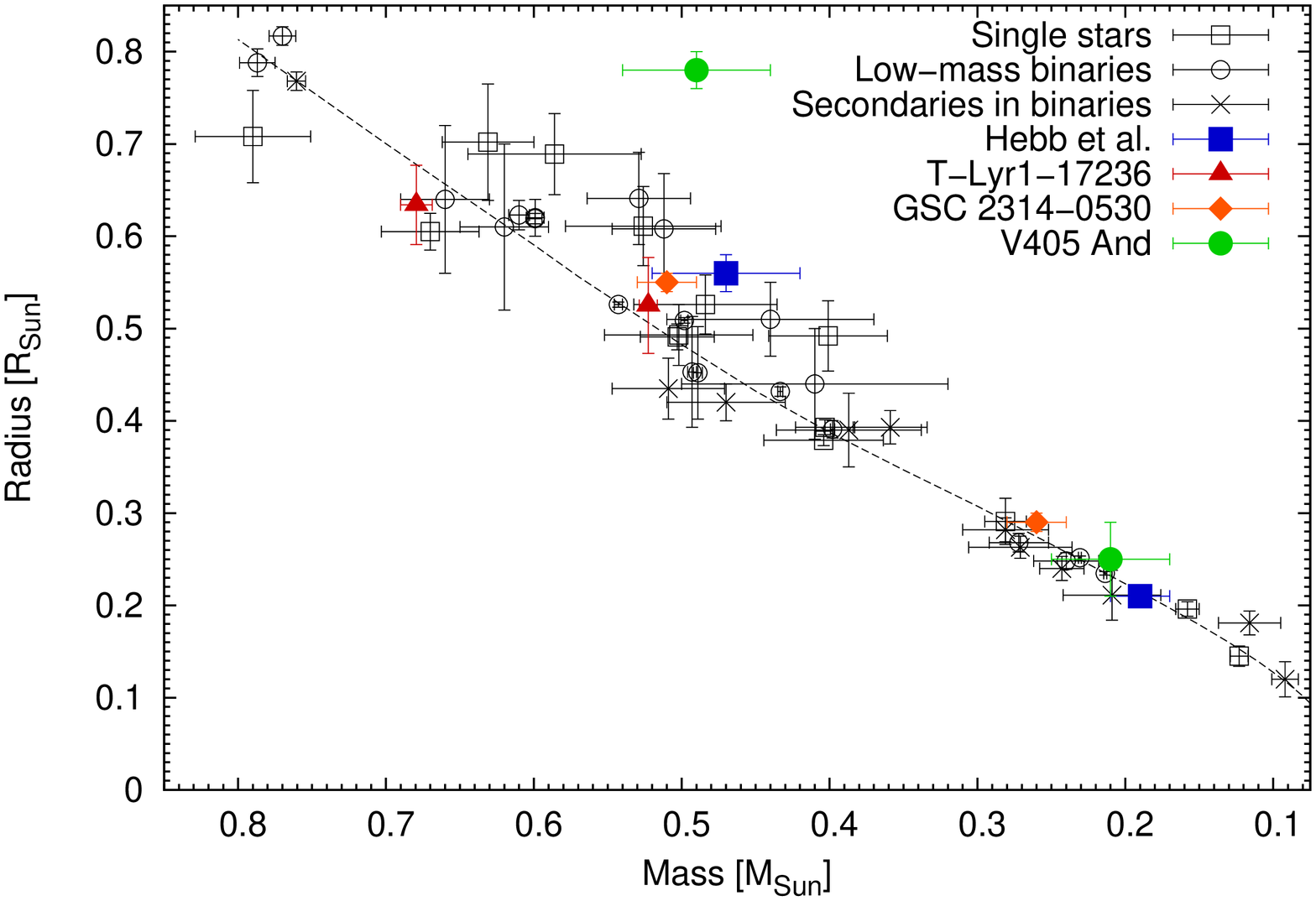} 
\caption{
Left: Fits to $BV(RI)_C$ light curves of V405\,And. The first column shows the results from \cite{vand}, the rest show new results.
Right: Mass-radius diagram for 5\,Gyr stars from \cite{baraffe} (with continuous line). Dots show measurements of \cite{MR1}, \cite{MR2} and \cite{MR3}. Filled symbols denote V405\,And, GSC\,2314-0530 \citep{gsc}, T-Lyr-17236 \citep{tlyr}, and 2MASS 04463285+1901432, a binary in NGC 1647 \citep{hebb}.
}

   \label{fig1}
\end{center}
\end{figure}

%

We have obtained new $BV(RI)_C$ photometry with the 1\,m RCC telescope at Piszk\'estet\H{o} between JDs 2455148 and 2455531 (2009 November--2010 November, about 700 days after the  light curves modelled in \citealt{vand}). Previously we found the light curve to be stable \citep{vand}, but during the time of the new observations the surface seemed to evolve significantly. The main spotted area moved from phase $\approx 0.5$ to phase $\approx 1$ in the first two new seasons, and in the last season two active nests were observed: around phases 0.1 and 0.5.

Using the same modelling method described in \cite{vand}, we modelled the new observations using PHOEBE \citep{phoebe} and SpotModeL \citep{sml}. The light curves and the fits are plotted in Fig. \ref{fig1}. The models fitted to the new observations left the system parameters unchanged. This indicates, that the radius of the primary is indeed much larger than expected. The two known similar binaries with similar structure, GSC 2314-0530 \citep{gsc} and the one from \cite{hebb} does not show this behaviour, although the primary of 2MASS 04463285+1901432 \citep{hebb} has somewhat larger radius. This indicates that V405\,And is a currently unique system, definitely worth for further studies.

\acknowledgements
The authors acknowledge support from the Hungarian Research Grant OTKA K-081421 and the "Lend\"ulet" Program of the Hungarian Academy of Sciences.


\end{document}